\documentstyle[pre,aps,epsfig,twocolumn]{revtex}
\def\break#1{\pagebreak \vspace*{#1}}

\begin{document}

\addtolength {\oddsidemargin} {-0.9cm}
\addtolength {\topmargin} {1cm}
\setlength{\parindent}{0.4cm}

\title{Stochastic Maps, Wealth Distribution in 
Random Asset Exchange Models and the Marginal Utility of Relative Wealth}

\author{Sitabhra Sinha}

\address{The Institute of Mathematical Sciences, C. I. T. Campus,
Taramani, Chennai - 600 113, India.}

\maketitle

\widetext

\begin{abstract}
We look at how asset exchange models can be mapped to random iterated
function systems (IFS) giving new insights into the dynamics of wealth
accumulation in such models. In particular, we focus on the ``yard-sale"
(winner gets a random fraction of the poorer players wealth) and the
``theft-and-fraud" (winner gets a random fraction of the loser's
wealth) asset exchange models. Several special cases including 2-player
and 3-player versions of these `games' allow us to connect the results
with observed features in real economies, e.g., lock-in (positive
feedback), etc. We then implement the realistic notion that a richer agent
is less likely to be aggressive when bargaining over a small amount with a
poorer player. When this simple feature is added to the yard-sale model,
in addition to the accumulation of the total wealth by a single agent
(``condensation"), we can see exponential and power-law distributions of
wealth. Simulation results suggest that the power-law distribution occurs
at the cross-over of the system from exponential phase to the condensate 
phase.
\end{abstract}

\vspace{0.5cm}

\narrowtext

\section{Introduction}
Recently, there has been a considerable amount of work done in developing
the statistical mechanics of economic activities leading to wealth accumulation
and distribution in society \cite{Hay02}.
One of the simplest class of models which explore these mechanisms are
the ``asset exchange models'' \cite{Isp98,Dra00,Cha00,Cha02,Cha03a,Cha03b}.
In analogy with the physics of ideal gases, economic agents can be viewed
as particles which have random elastic collisions with each other,
resulting in wealth circulation throughout the system. One reliable
indicator of whether these models reflect economic reality is to test
whether they reproduce the observed wealth distribution in various 
societies.

It has been known now for over a century that almost all human societies
tend to exhibit the same type of wealth distribution. If $P ( x )$ is the
probability distribution for income or wealth $x$ of individuals, then
for large $x$, it follows the so-called Pareto law:
$$ P ( x ) \sim x^{- ( 1 + \nu)},$$
i.e., a power law distribution with the exponent $\nu$ between 1 and 2
\cite{Lev97}, while for small $x$, an exponential distribution is 
observed \cite{Dra01,Sou02}.

Unfortunately, the simplest asset exchange models do not show
this distribution. The asymptotic states of these models exhibit
either an exponential phase, or even more extremely, all the wealth
condensing into the hands of a single individual; power-law
distributions, if seen at all, turn out to be 
transient \cite{Isp98,Dra00,Cha00,Cha02}. However, recent work on 
the effects of introducing random 
saving propensities in the asset exchange models, have shown asymptotic 
power law distribution similar to those observed in 
reality \cite{Cha03a,Cha03b}. 
In this paper, we look at another possible modification of the
asset exchange 
\break{1.85in}
model: the same amount of money may 
have different relative values to a rich agent and a poor agent. In other
words, the relative importance of making a net gain in a round of trading
is dependent on the relative wealths of the agents involved. By introducing
this simple principle into the model, we observe a wide range of 
distributions: from an exponential phase to a condensate phase, with a
power-law distribution appearing in the transition region between the
two phases. 

We consider a simple model of a closed economic system where the total
wealth (amount of money) available for exchange, $M$, and the total number
of agents, $N$, trading with each other, are fixed. Wealth is neither created
nor destroyed, but only change hands through trading between agents.
Further, the system is observed only at discrete time intervals
$t = 0, 1, 2, 3, ...$.
Each agent $i$ has some wealth $x_i (t)$ associated with it at some
time step $t$. Starting from an arbitrary initial distribution of wealth
($x_i (0)$, $i = 1, 2, 3, ....$), during each time step two randomly chosen 
agents $i$ and $j$ 
exchange a fraction of their combined wealth. Each such transaction is
obeying the constraint that the combined wealth of the two agents
is conserved by the trade, and that neither of the two has
negative wealth after the trade (i.e., debt is not allowed). 
In general, one of the
players will gain and the other player will lose as a result of the trade.

Different types of exchange models are defined based on the choice of the 
fraction
of wealth that will be exchanged in a trade. If the wealth exchanged is a
fraction of the wealth owned by the poorer of the two agents, 
then the model (in accordance with the terminology introduced in
Ref. \cite{Hay02}) is the
so-called ``Yard-Sale'' model (YS), whereas if the exchanged amount is a
fraction of the losing agent's wealth, it is the ``Theft-and-Fraud'' model
(TF) \cite{Hay02}. These names reflect the fact that usually (e.g., in a 
yard sale) the richer agent is unlikely to stake its entire holdings in a
trade with a poorer agent. The only circumstances during which such an
event is likely to occur is when the poorer agent is dishonest, and either
the exchange itself, or the wealth of the poorer agent, is unknown to the
richer agent (corresponding to theft and fraud respectively).

If we consider an arbitrarily chosen pair of agents ($i$, $j$) who trade
at a time step $t$ resulting in a net gain of wealth by agent $i$, 
then the change in their wealth as a result of trading is:
\begin{equation}
x_i (t + 1) = x_i (t) + \Delta x; ~~~ x_j (t + 1) = x_j (t) - \Delta x,
\end{equation}
where, $\Delta x$ is the net wealth exchanged between the two agents. 
In the YS model
\begin{equation}
\Delta x = \alpha ~min ( x_i(t), x_j (t) ),
\end{equation}
while in the TF model 
\begin{equation}
\Delta x = \alpha ~x_j (t) ~~~~{\rm (agent} ~j {\rm ~has ~lost)},
\end{equation}
with $\alpha$ as a uniformly distributed random number in the 
interval $[0, 1]$.
Whether agent $i$ or $j$ will `win' in a particular trading encounter
is decided by the toss of a fair coin, i.e., each has a probability
1/2 of making a net gain. A possible variant, where,
the two agents randomly redistribute their total wealth can be 
seen as a manifestation of the TF model.

In the next section, the asset exchange models are seen as a class of
random dynamical systems. This picture allows us to understand in 
simple terms
various features of the distribution seen in the two models. Section 3
introduces the concept of diminishing bargaining efficiency as the
wealth of an agent is increased. Results of 2-agent and $N$-agent asset
exchange models are given. Finally, we conclude with a summary and 
discussion of possible directions for future work.

\section{Asset Exchange Models as Stochastic IFS}
If we consider $\alpha$ to be a constant in Eqs. (2) and (3), then
for $N$ = 2, the asset exchange model (YS or TF) is a system of two maps 
of the
unit interval [0,1] onto itself, with the system randomly switching
between the two maps. The map selected at a particular instant depends 
upon which agent wins in that particular trading round.
Such stochastic dynamical systems are called Iterated Function 
Systems (IFS) \cite{Bar93}. 
The 2 map IFS corresponding to the YS and TF models are shown in
Fig. 1 for $\alpha$ = 0.5. 

\subsection{2-agent models}
Let us consider the YS model with $N$ = 2, and total 
wealth $M = \Sigma_{i=1}^N
x$ =1 (i.e., normalized). Then, the state of such an economy at any given
time $t$, is completely specified by the wealth of any one of the 
agents, $x (t)$ (since the other agent's wealth is $1 - x(t)$). For
constant $\alpha$, the corresponding IFS is given by
\begin{equation}
{\rm Map ~1:~} x (t+1) = (1 + \alpha) ~x (t), ~{\rm if} ~x (t) < 0.5,
\end{equation}
$$~= x (t) + \alpha (1 - x (t)), ~{\rm otherwise},$$
and,
\begin{equation}
{\rm Map ~2:~} x (t+1) = (1 - \alpha) ~x (t), ~{\rm if} ~x (t) < 0.5,
\end{equation}
$$~= x (t) - \alpha (1 - x (t)), ~{\rm otherwise}.$$
For $\alpha = 0$, the initial distribution is unchanged by the IFS,
but for any $\alpha > 0$, the final distribution corresponds to two 
delta function peaks at $x = 0$ and $x = 1$. In other words, the entire wealth
eventually ends up in the hands of one of the two agents through a process
of gradual wealth condensation. The transition
to this {\em condensate phase} from an arbitrary initial 
distribution takes longer
and longer time as $\alpha \rightarrow 0$, in a process analogous to
{\em critical slowing down}. 

In the TF model with $N$ = 2, the wealth dynamics is given by the IFS:
\begin{equation}
{\rm Map ~1:~} x (t+1) = (1 - \alpha) ~x (t) + \alpha,
\end{equation}
and,
\begin{equation}
{\rm Map ~2:~} x (t+1) = (1 - \alpha) ~x (t).
\end{equation}
The asymptotic state for a constant value of $\alpha (> 0.5)$ 
is a Cantor set fractal 
distribution. In particular, for $\alpha$ = 2/3, the final distribution is
the middle-third Cantor set, generated by successively dividing intervals
into three equal parts and then removing the central section. Therefore,
the wealth possessed by an agent at any given time form a discontinuous 
range of values.

For randomly varying $\alpha$, the asymptotic distribution of TF model
is composed
of an infinite number of Cantor sets generated by the different values of
$\alpha$. This turns out to be a power law distribution with an
exponent $\simeq 0.45$. The occurrence of a power-law or 
scale-free distribution can be understood as follows. Each of the Cantor
set distributions generated for a fixed $\alpha$ has a length scale
associated with it, corresponding to the fraction of an interval
removed recursively during its generation. For random $\alpha$, since
all Cantor sets are represented, this implies that all length scales are
present in the asymptotic distribution. This results in a scale-free
distribution for randomly varying $\alpha$.

To understand why YS and TF models lead to very different asymptotic
distributions, we look at the effect of a sequence of unfavorable outcomes
(i.e., losses) on the wealth of the richer agent (Fig. 2). Let us suppose
that in both models, the rich agent (A, say) owns a significant fraction of the
total wealth ($x = 0.95$, say). If we now look at the result of a series
of losses, we find that agent A is much less affected in the YS model
than in the TF model, and the larger the initial wealth of A, the greater
is the difference between the two models. This means that, if initially
one of the agents acquire a significant fraction of the total wealth (by
random fluctuations), then the dynamics of the YS model assures that the agent
will consolidate this position. The greater the value of wealth acquired
by the agent, the larger is the number of successive unfavorable outcomes
needed (and therefore, more and more unlikely to occur) 
to shift it from the status of the richer agent.

This observation is further strengthened by observing the $n$-th return 
maps for each of the two systems (Fig. 3). 
For the YS model, all the higher iterate maps have
{\em stable} fixed points only at $x = 0$ and $x = 1$. The TF model on the
other hand have increasing number of stable fixed points distributed
over a wide range as the order of the return map is increased. This
immediately implies that randomly switching between maps in the TF
model will lead to a fairer distribution of wealth, whereas that will
not be the case in the YS model. Further, for YS model we find that
the attractor where the system will eventually end up in, is decided
by the first few outcomes of bargaining among the two agents. Once
the system enters the basin of attraction of one of the attractors
($x = 0$ or $x = 1$) through a series of favorable outcomes, 
it takes many more unfavorable outcomes for it to come out and
enter the other attractor's domain. If
one of the agents, by chance, wins the first few successive bargaining
rounds then it is almost bound to be the eventual winner. This is
reminiscent of the phenomena of positive feedback leading to 
`lock-in' in real economic systems \cite{Art89}. 
\subsection{3-agent models}
The 3-agent TF model (with the total wealth normalized to 1) can be 
represented in the triangular region [(0,0), (0,1), (1,0)] in the
two-dimensional plane. This is because we need to only explicitly specify
the wealths $x_1~(t)$ and $x_2~(t)$ of two agents (A and B, say), 
the third agent (C) having wealth $1 - x_1~(t) - x_2~(t)$.

The 3-agent model is the simplest case where we can study the effect
of different types of interaction coupling among agents. For example,
instead of allowing all the agents to trade among themselves, we can
forbid trade between agents B and C (say). Therefore, B and C can only
trade with each other through a ``go-between" (in this case, agent A).
Our simulations showed no differences in the results for the two schemes.
Fig. 4 (a) shows the attractors corresponding to the two types of
coupling among agents.
The asymptotic distribution is again a power law, with the exponent
value of $0.5 \pm 0.005$ (Fig. 4 (b))
for both types of network interaction structure.

\subsection{$N$-agent models}
As the number of agents $N$ is increased, the models become increasingly
difficult to understand in terms of dynamical systems. However, most
of the features of the 2- and 3-agent games carry over to the $N$-agent
case. For example, the observation that all wealth condenses into the
hands of a single agent, holds true in the $N$-agent YS model as $N
\rightarrow \infty$. In the TF model, as $N$ increases, the asymptotic
wealth distribution becomes exponential. 

A few special cases can be understood completely even in the $N
\rightarrow \infty$ limit. One of these is the case when $\alpha = 1$. 
For the YS model, this corresponds
to a ``Double-or-nothing" scenario, where the poorer agent stands to either
double its wealth (if it wins) or lose everything to the richer agent (if
it loses). The richer agent, on the other hand, stands to lose or gain
only an amount of wealth equal to that owned currently by the poorer agent.
The corresponding situation in
the TF model gives the winner (independent of whether the richer or
the poorer agent wins) all the assets of the loser. Therefore,
this situation corresponds to a ``Winner-take-all" scenario. It can be
easily seen that both cases quickly lead to the concentration of wealth 
into the hands of a steadily decreasing minority.

\section{Diminishing Bargaining Efficiency with Wealth}
We have so far assumed that the probability that an agent will gain net
wealth is independent of its wealth. However, in any real situation, it
is unlikely that an agent who owns 1,000 units of wealth (say) will be
as concerned about winning or losing 1 additional unit, as 
an agent who has only 1 unit. Therefore, the relative value of the 
amount of wealth won or lost by an agent is clearly a function of
its wealth at that given time. This results in the increasing
aggressiveness of the poorer agent in getting a favorable deal during
any trade with a wealthier agent.

This is implemented in the asset exchange model by expressing the
probability that agent $i$ will win when trading with 
agent $j$, $p (i | i, j)$, as a `Fermi function':
\begin{equation}
p (i | i, j) = \frac{1}{1 + exp (\beta [\frac{x_i}{x_j} -1])},
\end{equation}
with $\beta$ parametrising the significance of the relative value
of wealth between trading agents. For $\beta$ = 0, the original
asset exchange model is retrieved, where the probability that any agent
wins a round of trading is 1/2, independent of their wealth.
When $\beta > 0$, the poorer agent has a higher probability of
winning, the difference from the original probability ($= 1/2$) depending
on the value of $\beta$. In the special circumstance when the two agents 
have the same wealth, the probability of each of them winning is 1/2, 
irrespective of the value of $\beta$. 

As $\beta \rightarrow \infty$, it becomes certain that in any encounter
the poorer agent will win. Intuitively it is clear that this will ensure a 
fairer distribution of wealth. In fact, when $\alpha$ is a constant, 
the corresponding stochastic IFS reduces to a deterministic map; when
$N$ = 2, it is a map of the unit interval [0,1] onto itself (Fig. 5).

In the YS model, with $\alpha$ = 1 (``Double-or-nothing'' for the
poorer agent), 
the wealth distribution
remains in the condensate phase even as $\beta \rightarrow \infty$. This is
because, although wealth changes hands frequently, the number of solvent
agents who can trade steadily decreases over time. Wealth gets accumulated
into the hands of a steadily diminishing number of agents, 
although the label of the wealthiest agent keeps changing.
But for any $\alpha < 1$, the wealth distribution will tend to
be fairer, i.e., the asymptotic distribution no longer corresponds to all the
wealth ending up with just one agent. The chaotic map corresponding to
the YS model at $\beta \rightarrow \infty$ (for a fixed value of $\alpha$) 
ensures a more uniform distribution of wealth among the agents.

When we implement this principle in the 2-agent YS model with randomly 
varying 
$\alpha$, we immediately find that varying $\beta$ completely alters
the asymptotic wealth distribution. At the limit $\beta \rightarrow \infty$
the distribution function is triangular or `tent'-shaped:
$$ P (x) = 4 x, ~{\rm if}~ x \leq 0.5;~= 4 (1-x), ~{\rm otherwise}, $$
where $x$ is confined to the unit interval. As $\beta$ is decreased,
we find that the central peak of the distribution (at $x = 0.5$) gradually
diminishes, while the tails of the distribution (at $x = 0$ and $1$) 
gradually start rising (Fig. 6).
The distribution finally becomes identical to the delta function
peak distribution of the conventional YS model as $\beta \rightarrow 0$.

In the 2-agent TF model (Fig. 5 (right)), if $\alpha$ is fixed, then at 
$\beta \rightarrow \infty$, the asymptotic state
is a 2-cycle, with the two players alternately switching between
two wealth values $x_1, x_2$ ($x_1 > 0.5 > x_2$). The exact numerical
value of $x_1, x_2$ depends on $\alpha$. With randomly varying $\alpha$
we find a smooth asymptotic distribution which peaks at the center ($x = 0.5$).
Unlike the TF model, here the distribution is not piecewise linear.
With decreasing $\beta$, the central peak diminishes with corresponding
increase at the tails of the distribution (Fig. 7).

When we implement the principle in the $N$-agent YS model, with $\alpha$ fixed
to a constant value ($= 0.5$, say),
we find that a power-law distribution is observed at intermediate values
of $\beta$. The same result is seen when $\alpha$ is randomly
varied (Fig. 8). 
For large value of $\beta$, the asymptotic distribution is exponential
for large wealth (Fig. 8 (a)), with an initial power law increase (which
disappears with decreasing $\beta$). With increasing $\beta$, the distribution
shows an increasing peak at the most probable value of $x = 1$, i.e., the
average wealth per agent. In other words, as the effect of the relative
value of wealth becomes more pronounced, it becomes more likely
that every agent will have the same amount of money (on average).
As already mentioned, at intermediate values of $\beta$,
a power-law distribution is observed (Fig. 8 (b)).
As $\beta$ is decreased further, a condensation at very high value of 
wealth is observed in addition to the power law distribution. 
Further decrease of $\beta$ leads 
to the increasing condensation of the wealth in the hands of fewer and 
fewer agents, until, at $\beta = 0$,
a single agent acquires all the wealth (the conventional YS model).

Note that if we implement {\em increasing} bargaining efficiency with
increasing wealth, we have a situation essentially similar to the
`greedy exchange' model investigated in Ref. \cite{Isp98}. This
possibly corresponds to the condition that the wealthier agent has greater 
chance of dictating terms to the poorer agent during any bargaining.

\section{Discussion}
In this paper we have first discussed random asset exchange models
as a class of stochastic dynamical systems. This allows us to
understand why different asymptotic distribution are observed in
the YS and TF models. We can also connect the features observed in
these models with phenomena like positive feedback (`lock-in') seen
in real economic situations. We have then introduced the realistic
notion that the `value' of a certain quantity of wealth is relative 
to the wealth currently
owned by an agent, i.e., to a very wealthy agent the gain or loss of 
a very small quantity is not as important as it is to a very poor agent.
Implementing this principle, we observe exponential distribution of
wealth among agents, in addition to the
condensate phase observed in the original YS model. More interestingly, 
we see a power-law distribution of wealth
in the region of transition between the exponential and condensate phases.

The use of a Fermi function (Eq.~(8)) to express the probability of an agent to 
win a particular exchange with another agent, $p (i|i,j)$, as a function of 
their relative wealth, $x_i / x_j$, can
be interpreted as an implementation of a principle of {\em diminishing
marginal utility of relative wealth}. If the utility $U$ is expressed as a 
function of the relative wealth $w$, then the probability $p (i|i,j)$ is 
related to the marginal utility $\partial U/\partial w$. Therefore, the
probability of winning, and hence,
the marginal utility corresponding to an utility function which initially 
rises rapidly with $w$, but for higher values of $w$ shows very little 
or no growth, is represented very well by a Fermi function. In the
$\beta \rightarrow \infty$ limit, Eq.~(8) implies that the utility function 
increases linearly upto some maximum value $U = U_{max}$ (when the relative
wealth $w = 1$), and then stays constant at all higher values of $w$.
On the other hand, when $\beta = 0$, the utility function is a strictly
linear increasing function of the relative wealth.

A possible interesting aspect not fully explored is the role of the network
structure of interaction among the agents. There has been some study
of small-world effects in wealth distribution through interactive
multiplicative stochastic process \cite{Sou01}. In the conventional
YS model, the introduction of small-world or regular structure, requiring
agents to interact preferentially with some agents, does not change
the eventual delta-function peak distribution of wealth. The only effect of a
regular network structure is that multiple delta-function peaks coexist,
depending upon the allowed range of interaction. We have not carried out
a corresponding study of the model under the condition of diminishing
bargaining efficiency with wealth.

Another possibly fruitful area of future work is the connection of
the models studied in this paper to mass-aggregation models allowing
diffusion, aggregation and dissociation \cite{Maj98}. Such models show
nonequilibrium phase transitions, from a state where the steady state
mass distribution decays exponentially to another state where an
infinite aggregate is observed in addition to a power law distribution.
The principal difference with the models that we study here is that
in mass aggregation all the units exchanged have the same fixed mass value, 
whereas here we have no limit to
the smallest amount of wealth an agent can possess and exchange. 
However, it is striking that both types of models show similar phases.

\vspace{0.2cm}
I thank  Bikas K. Chakrabarti and Maitreesh Ghatak for helpful discussions.
This work was partly inspired by the article of Brian Hayes 
(Ref. \cite{Hay02}).

\vspace{2cm}
\begin{center}
\bf{FIGURES}
\end{center}

\begin{figure}[t!]
\centerline{\includegraphics[width=0.95\linewidth,clip] {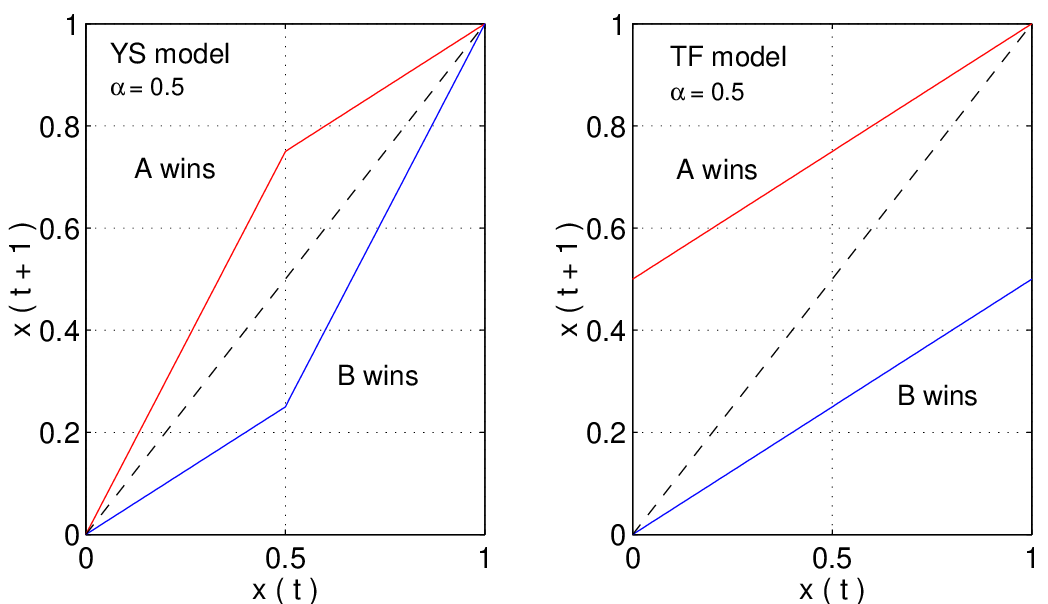}}
\vspace{0.5cm}
\caption{\small The IFS corresponding to the 2-agent (left) YS model and 
(right) TF model with $\alpha = 0.5$. The system stochastically switches
from one map to the other depending on which of the two agents (A, B)
wins a particular round of trading.}
\end{figure}

\begin{figure}[t!]
\centerline{\includegraphics[width=0.95\linewidth,clip] {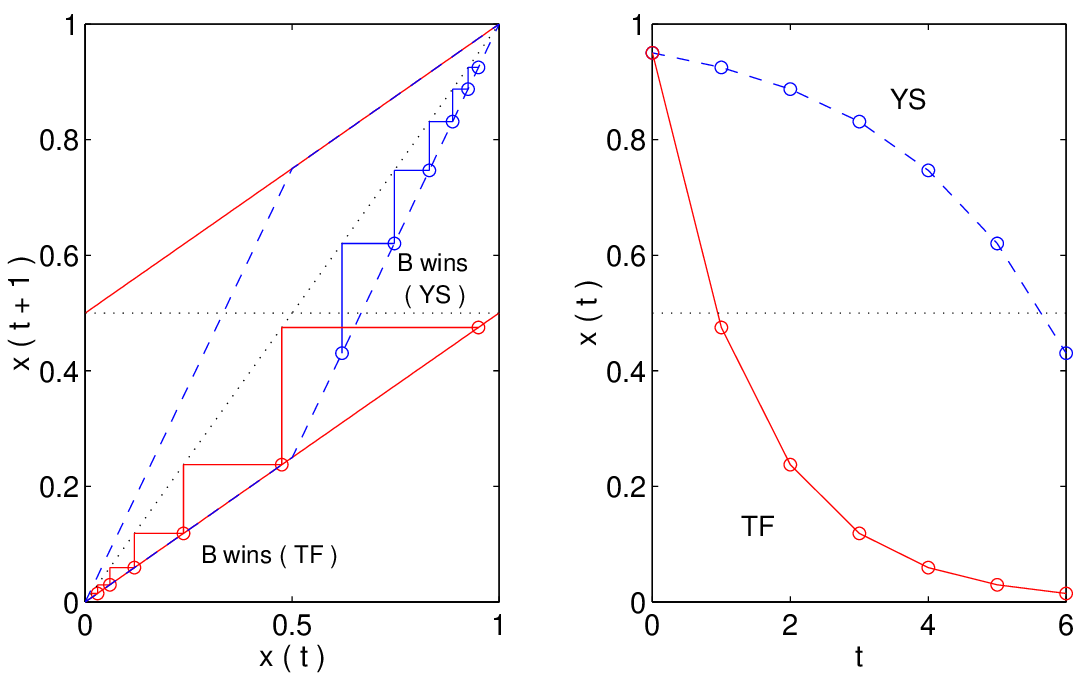}}
\vspace{0.5cm}
\caption{\small Difference between the 2-agent
YS model (broken lines) and TF model
(solid lines) in the effect
of a sequence of unfavorable outcomes on the wealth of the initially
richer agent (A) with $\alpha = 0.5$. 
In both models, agent A starts out with wealth $x_0 = 0.95$.
The return maps (left) of the two models show that the losses affect A 
more strongly in the TF model than in the YS model (especially when $x$ is 
large). The time evolution of the wealth fraction owned by A, $x (t)$ (right)
indicates that, while in the TF model, it takes only one loss to change the 
status of agent A from the richer ($x \geq 0.5$) to the poorer agent 
($x < 0.5$), in
the YS model, on the other hand, it takes {\em six} successive losses to
change the status of A. So, in the YS model, if initially one of the agents
acquire a significant fraction of the total wealth, this imbalance
will be consolidated in subsequent trading, as the occurrence of the 
number of unfavorable
outcomes needed to shift the balance in favor of the other agent becomes 
increasingly unlikely with increasing $x$.}
\end{figure}

\begin{figure}[t!]
\centerline{\includegraphics[width=0.95\linewidth,clip] {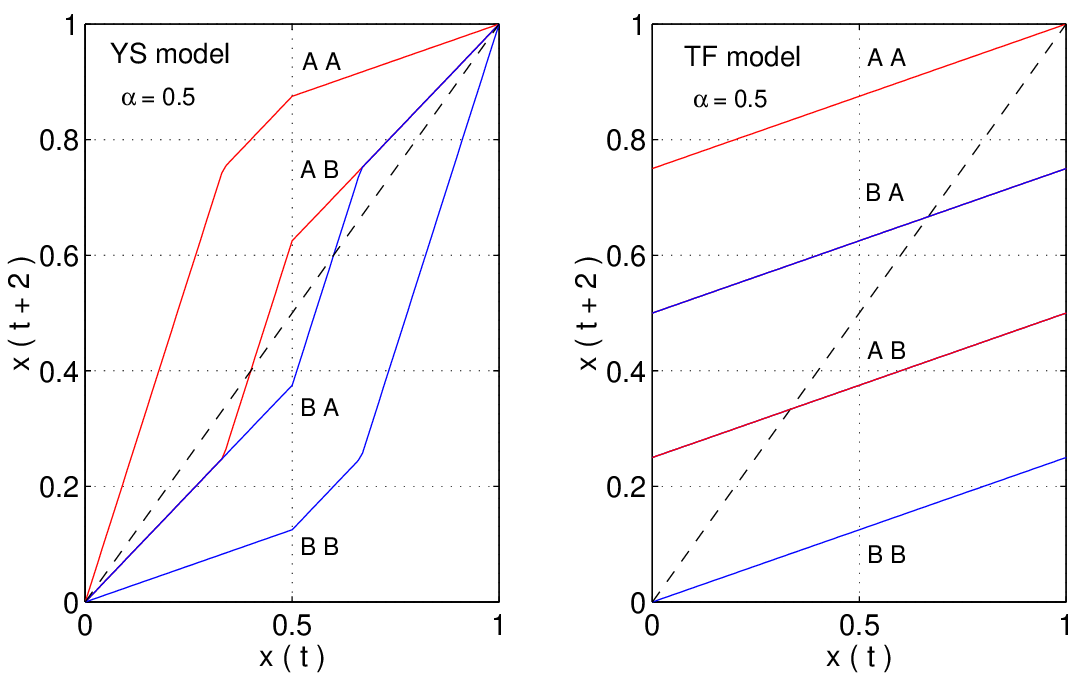}}
\vspace{0.5cm}
\caption{\small The second return map of the 2-agent (left) YS model and (right)
TF model with $\alpha = 0.5$. Both the models have four fixed points, but
while all four are stable in the TF model, only two ($x = 0$ and $x = 1$)
are stable in the YS model. Each curve is labelled by a pair of letters `XY',
indicating that it corresponds to the case where a win of agent X in the 
first time instant ($t + 1$), is followed by a win of agent Y in the next 
time instant ($t + 2$). Note that, in the YS model, for the case of both
agents winning once (i.e., AB or BA), the basin of attraction 
for $x = 1$ is larger than $x = 0$ when A wins first, and the reverse 
holds true when B wins first. This indicates the strong dependence of the final
state of the system on the initial states (i.e., the winner of the first
few `trades' has a very high probability of emerging as the eventual winner).}

\end{figure}

\pagebreak

\begin{figure}[t!]
\centerline{\includegraphics[width=0.8\linewidth,clip] {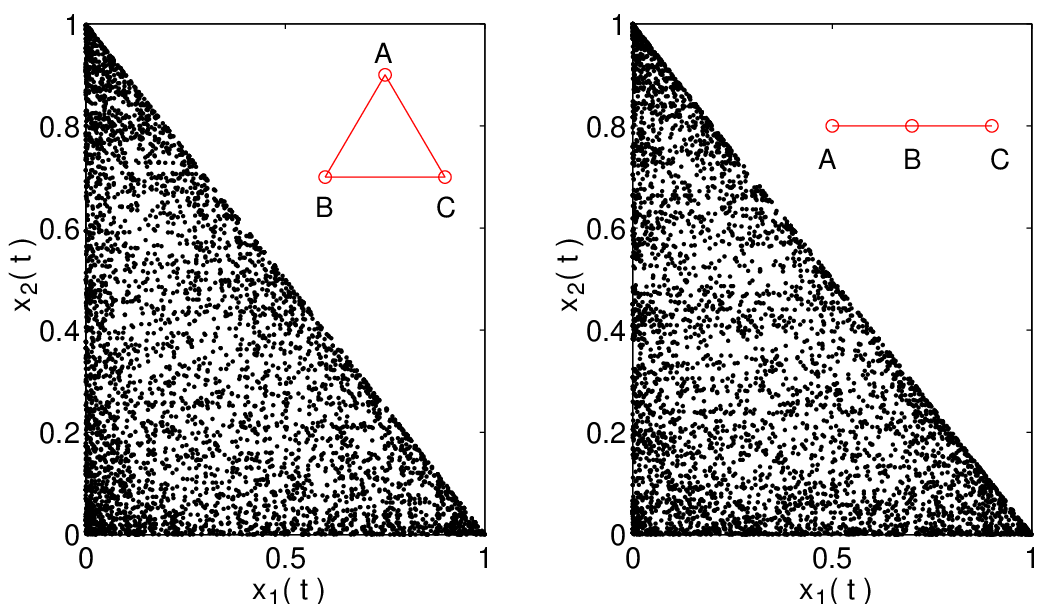}}
\centerline{( a )}
\vspace{0.5cm}
\centerline{\includegraphics[width=0.8\linewidth,clip] {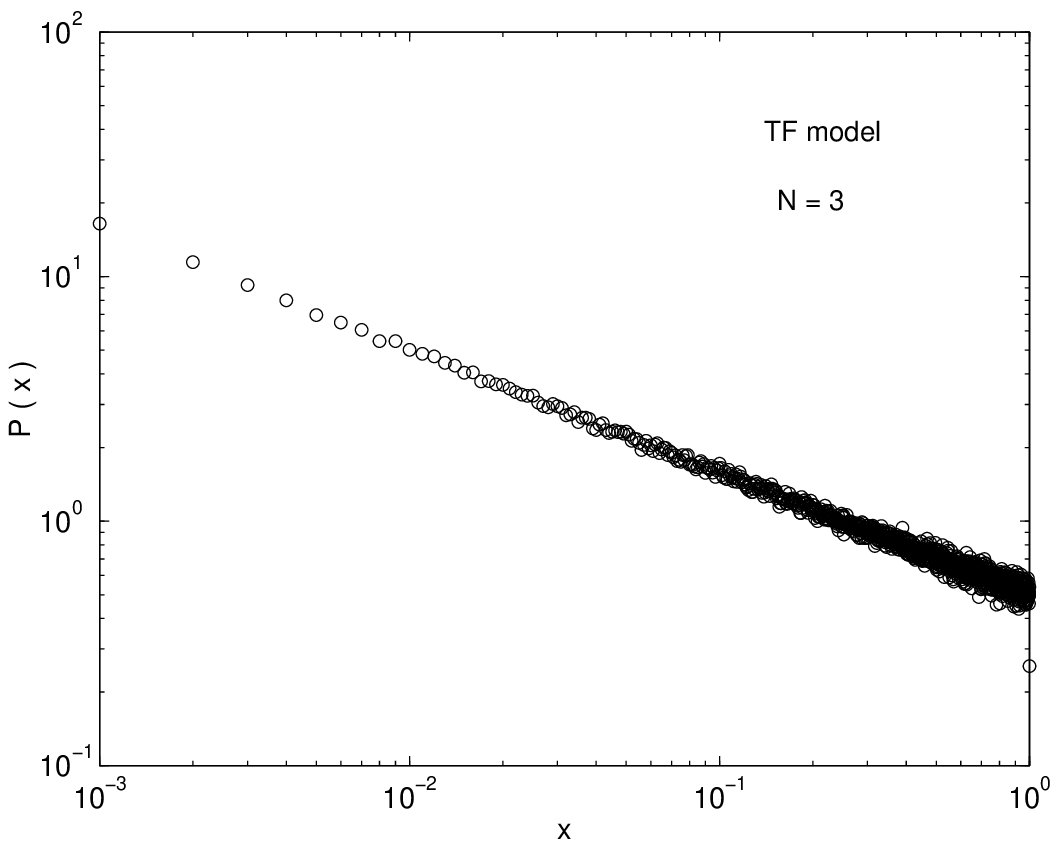}}
\centerline{( b )}
\vspace{0.5cm}
\caption{\small
(a) The attractors corresponding to the 3-agent TF model 
(randomly varying $\alpha$) for
the (left) unrestricted case, where all agents freely interact with each
other, and (right) the restricted case, where direct interaction between
two of the agents (B and C) is not allowed. (b) The asymptotic
probability distribution of wealth acquired by an agent in the
unrestricted case. The slope of the power-law distribution 
is $0.5 \pm 0.005$. The corresponding figure for the restricted case is
identical.}
\end{figure}

\begin{figure}[t!]
\centerline{\includegraphics[width=0.95\linewidth,clip] {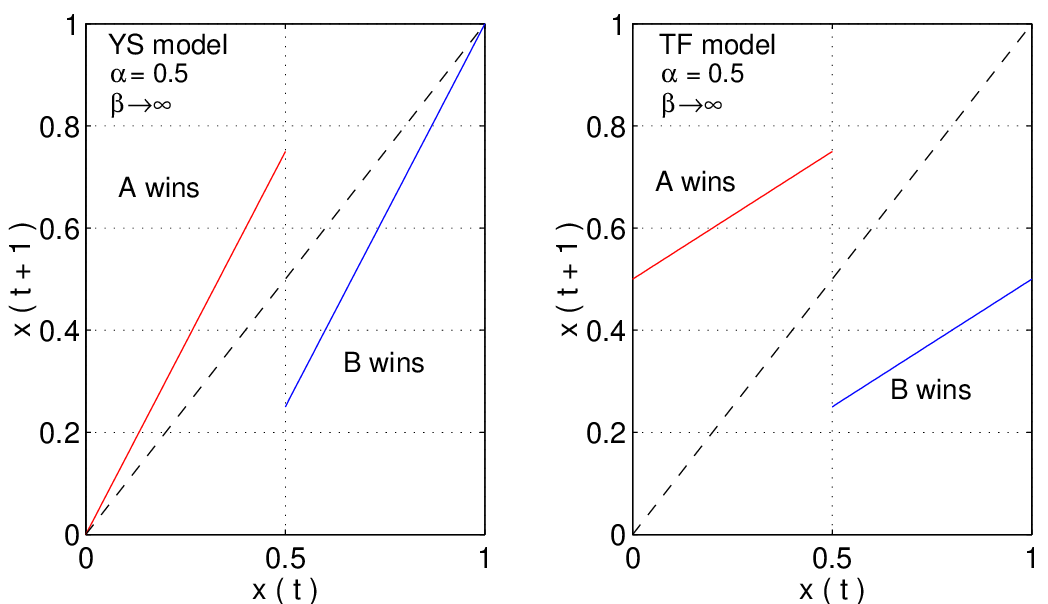}}
\vspace{0.5cm}
\caption{\small The deterministic maps corresponding to the 2-agent (left)
YS model and (right) TF model with $\alpha = 0.5$ and the relative
value parameter $\beta \rightarrow \infty$. In the YS model, the map
is chaotic, whereas, for the TF model the map dynamics converges to
a period-2 cycle.}
\end{figure}

\begin{figure}[t!]
\centerline{\includegraphics[width=0.95\linewidth,clip] {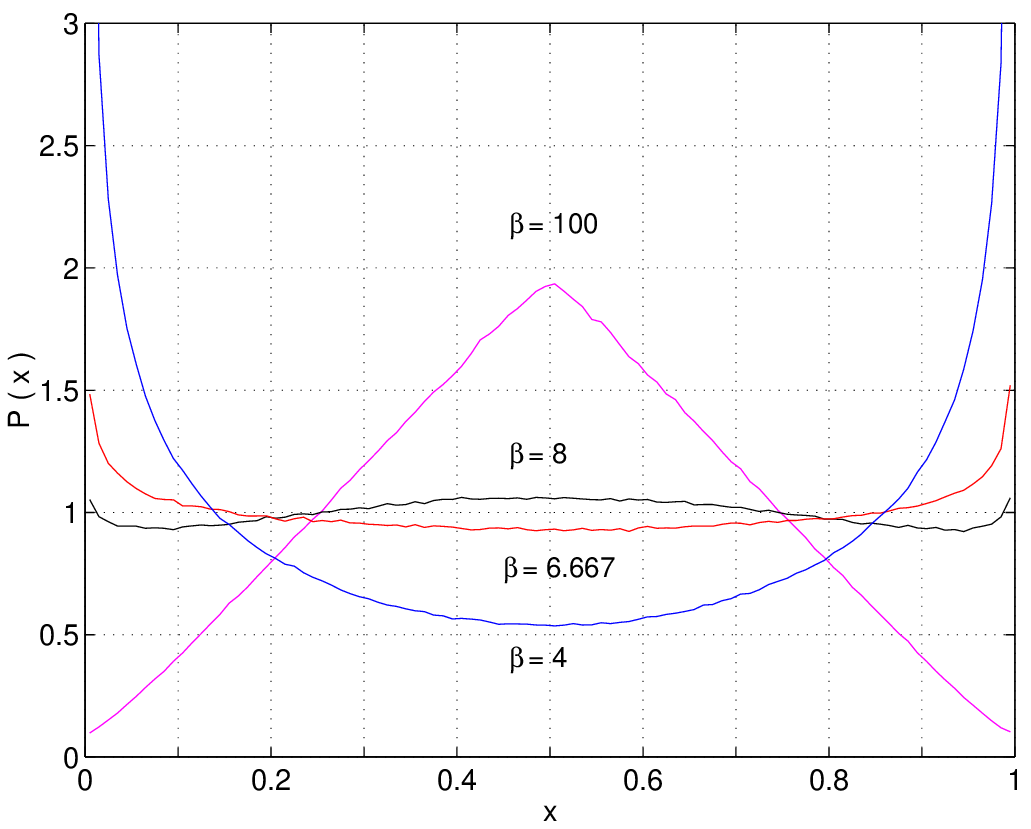}}
\vspace{0.5cm}
\caption{\small The asymptotic probability distribution of wealth
in the 2-agent YS model with randomly varying $\alpha$, as the
relative value parameter $\beta$ is gradually increased from
$\beta = 4$ to $\beta = 100$.}
\end{figure}

\begin{figure}[t!]
\centerline{\includegraphics[width=0.95\linewidth,clip] {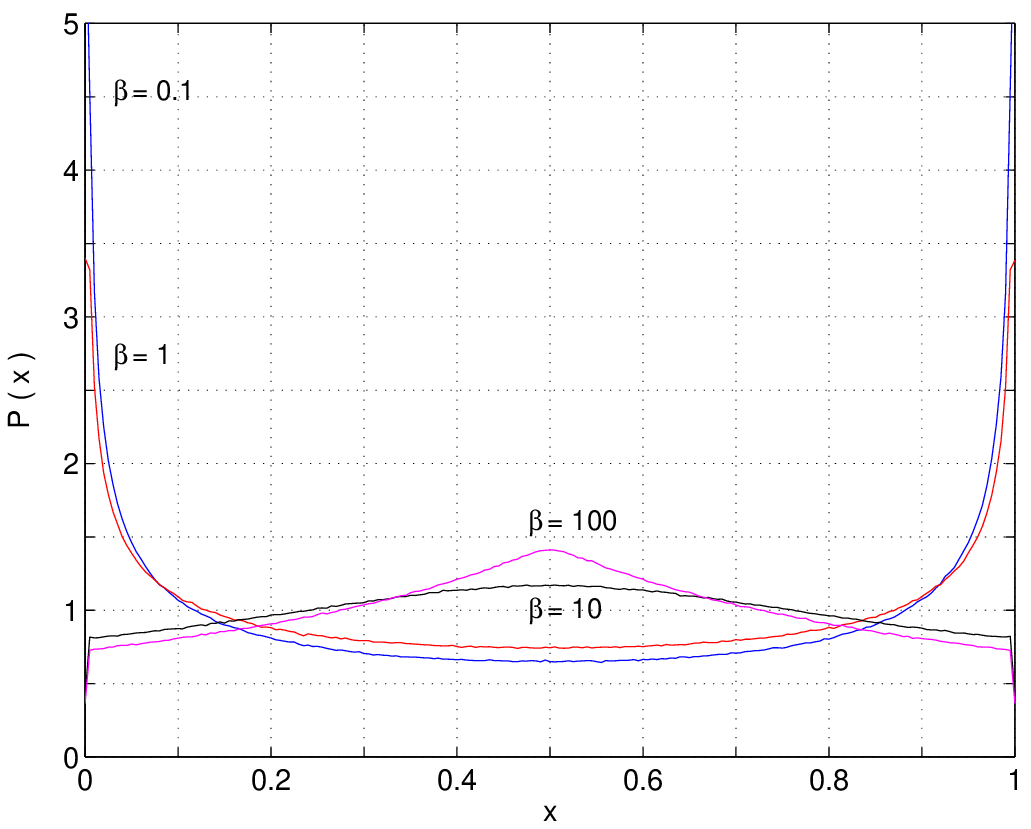}}
\vspace{0.5cm}
\caption{\small The asymptotic probability distribution of wealth
in the 2-agent TF model with randomly varying $\alpha$ as the
relative value parameter $\beta$ is gradually increased from
$\beta = 0.1$ to $\beta = 100$.}
\end{figure}

\begin{figure}[t!]
\centerline{\includegraphics[width=0.95\linewidth,clip] {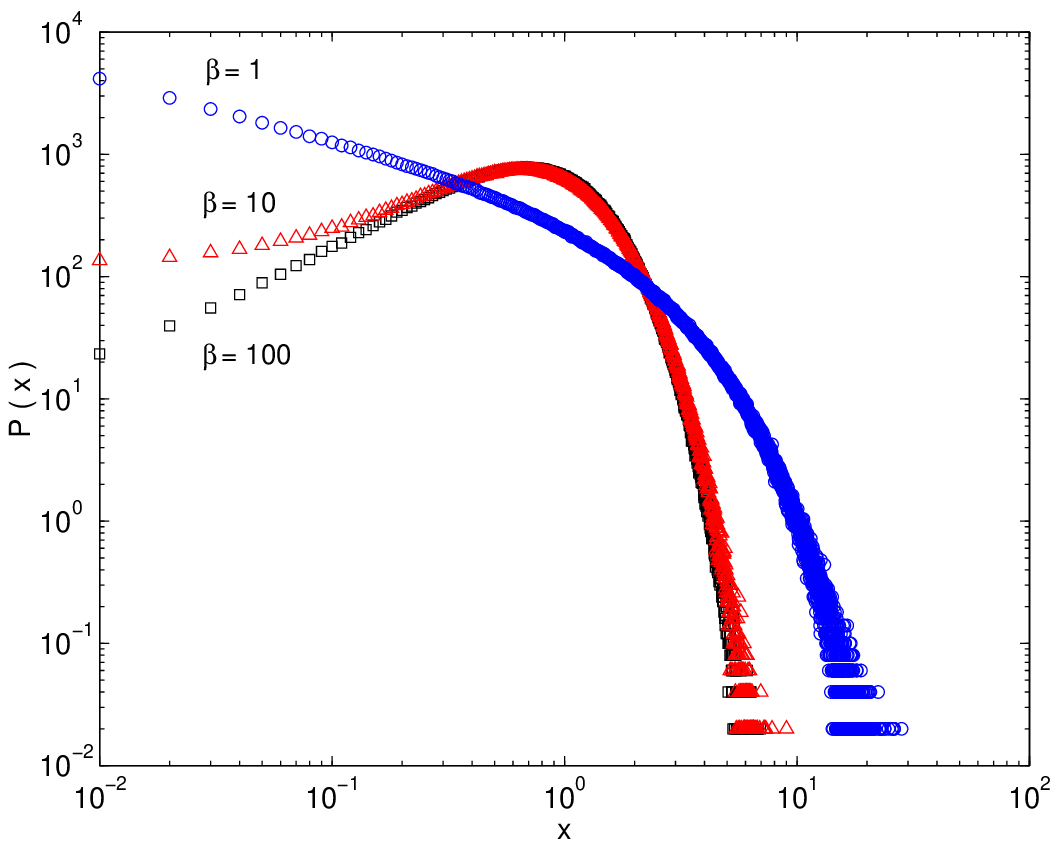}}
\centerline{( a )}
\vspace{0.5cm}
\centerline{\includegraphics[width=0.95\linewidth,clip] {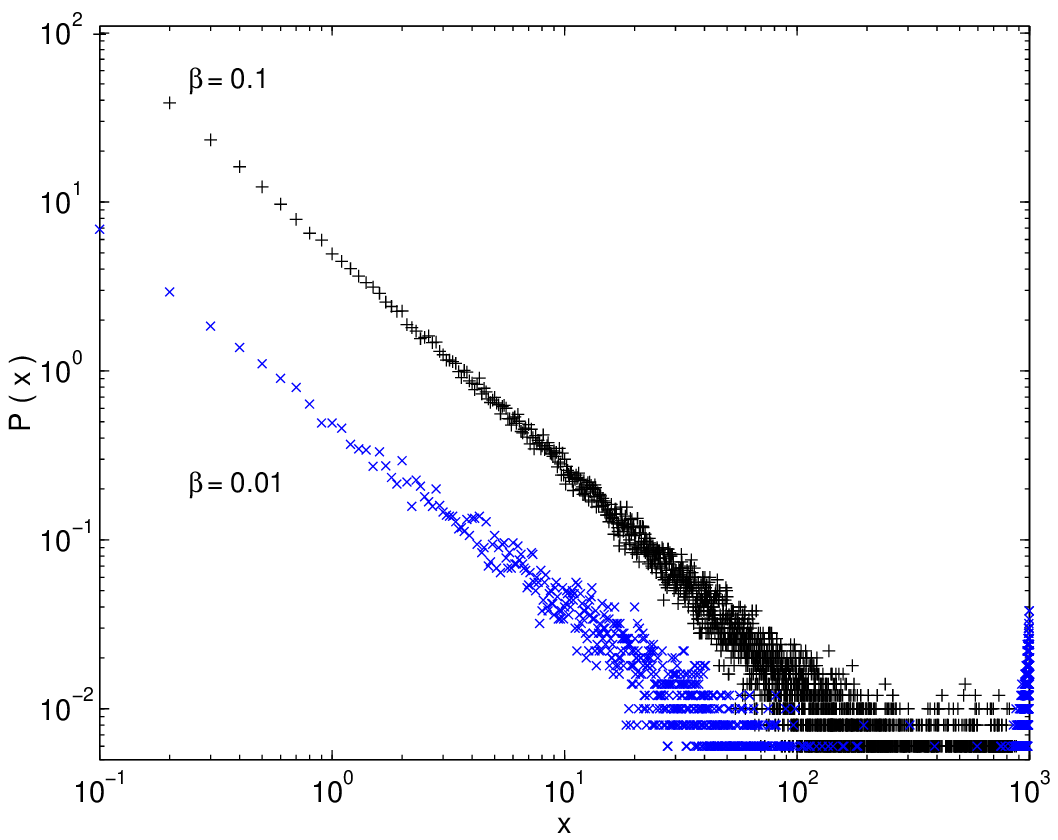}}
\centerline{( b )}
\vspace{0.5cm}
\caption{\small The probability distribution of wealth
in the YS model with $N = 1000$ at time $t = 1.5 \times 10^7$ 
($\alpha$ randomly varying) for
(a) $\beta = 1$ [circles], $= 10$ [triangles] and $= 100$ [squares] 
(exponentially decaying for large wealth), 
(b) $\beta = 0.1$ [pluses]( slope of the power-law curve is $1.30 \pm 0.05$)
and $\beta = 0.01$ [crosses] (slope of the power-law curve is $1.27 \pm 0.05$).
Note the condensation of wealth at $x \sim 10^3$ for $\beta = 0.01$.}
\end{figure}

\end{document}